\documentclass[superscriptaddress,reprint,aps,pre,amsfonts,amsmath]{revtex4-1}
\usepackage{inputenc}
\usepackage{amsmath}
\usepackage{amssymb, amsfonts}
\usepackage{graphicx,epstopdf,bm}
\usepackage[colorlinks,bookmarks,urlcolor=blue,citecolor=blue,linkcolor=blue]{hyperref}

\newcommand{\pd}[2]{\frac{\partial #1}{\partial #2}}
\newcommand{\pdd}[2]{\frac{\partial^{2} #1}{\partial #2 ^{2}}}
\newcommand{\pcd}[3]{\frac{\partial^{2} #1}{\partial #2 \partial #3}}

\newcommand{\abs}[1]{\left\vert #1 \right\vert}
\newcommand{\ave}[1]{\left \langle #1 \right \rangle}
\newcommand{\explr}[1]{\exp\left[ #1 \right]}

\newcommand{\chem}[2]{ {{#1\atop\longrightarrow}\atop{\longleftarrow\atop #2}} }
\newcommand{\bvec}[2]{\begin{bmatrix}#1\\#2 \end{bmatrix}}
\newcommand{\pr}{P}

\newcommand{\isd}{\rho} 
\newcommand{\pfac}{K}
\newcommand{\efn}[1]{\psi_{#1}}  
\newcommand{\efnref}[1]{\efn{#1}^{(r)}}
\newcommand{\efnabs}[1]{\efn{#1}^{(a)}}
\newcommand{\eigref}[1]{\lambda_{#1}^{(r)}} 
\newcommand{\eigabs}[1]{\lambda_{#1}^{(a)}} 
\begin{document}

\title{Bistable switching asymptotics for the self regulating gene}
\author{Jay Newby}
\email{newby.23@mbi.osu.edu}
\affiliation{Mathematical Bioscience Institute, Ohio State University, 1735 Neil Ave. Columbus, OH 43210}

\begin{abstract}
A simple stochastic model of a self regulating gene that displays bistable switching is analyzed.
While on, a gene transcribes mRNA at a constant rate.
Transcription factors can bind to the DNA and affect the gene's transcription rate.
Before an mRNA is degraded, it synthesizes protein, which in turn regulates gene activity by influencing the activity of transcription factors.
Protein is slowly removed from the system through degradation.
Depending on how the protein regulates gene activity, the protein concentration can exhibit noise induced bistable switching.
{ An asymptotic approximation of the mean switching rate is derived that includes the pre exponential factor, which improves upon a previously reported logarithmically accurate approximation.
With the improved accuracy, a uniformly accurate approximation of the stationary probability density, describing the gene, mRNA copy number, and protein concentration is also obtained.}
\end{abstract}
\maketitle

\numberwithin{equation}{section}
\renewcommand{\theequation}{\arabic{section}.\arabic{equation}}
\section{Introduction}

Metastability in a stochastic process is described by rare, noise-induced dynamical events.
For example, a Brownian particle in a double well potential, where the fluctuations are weak compared to the force of the potential, occationally jumps back and forth between each well.
Metastability is of particular interest in gene regulation circuits because rare extreme shifts in the expression of a gene can have a profound effect on the behavior of a cell \cite{eldar10a}.
The challenge for stochastic modeling is to elucidate possible metastable events and quantify the timescale on which those events are likely to occur.
Quantitative theoretical models can distinguish between events that may realistically occur on the timescale of cell division and those that occur on longer timescales.
Understanding the relative stability of metastable states in an artificial gene expression circuit is relevant in synthetic biology.
Because metastable events are by definition rare, an analysis based on direct simulation is computationally impractical.
In this paper, we derive an asymptotic approximation using perturbation theory.

One of the most difficult aspects of applying standard stochastic techniques to study gene regulation is accounting for reactions involving the gene.
Regulatory molecules, activators and repressors, bind to regulatory segments of DNA and interact with the gene promotor to affect the transcription rate (synthesis of mRNA).
There can be as few as one active copy of the gene in a given cell.
The case of linear feedback regulation is analytically tractable and many exact results are available \cite{hornos05a,visco08a,visco09a,venegas-ortiz11a}.
However, for the general case of nonlinear regulation, approximation methods are necessary.

Metastable behavior necessarily occurs under weak noise conditions, where fluctuations, whatever their source, are weak compared to deterministic forces.
A stochastic description of a given chemical reaction converges to deterministic mass action kinetics in the large system size limit where the number of molecules is large; this limit is sometime referred to as the large $N$ limit, where $N$ is the characteristic number of molecules.
Hence, it is natural to consider weak noise conditions for a stochastic chemical reaction to occur when $N$ is large but finite.
This is precisely the limit in which the chemical master equation is approximated by the chemical Fokker-Planck equation.

Clearly, no such limit is possible for a reaction involving a species having a single member.
However, if the reaction involving the gene is fast, one can obtain a deterministic description by taking an adiabatic limit, where the gene is described as switching between its various states infinitely fast so that it obtains an averaged transcription rate.
For example, a gene that switches between on and off states would, in the adiabatic limit, have an effective transcription rate scaled by the fraction of time spent in the on state.
A stochastic gene regulation model can then be said to be under weak noise conditions when it switches between its different states fast but not infinitely fast.

One could argue that mRNA should also be regarded as an adiabatic species.
In most situations {mRNA} copy number is quite low.
While {mRNA} are expensive to synthesize, a single copy is capable of producing many proteins.
If a gene expression model displays metastable behavior (i.e., weak noise conditions) and {mRNA} is present in small numbers, then it follows that the mRNA transcription and degradation must be fast (on the same time scale as promotor switching).

Methods for approximating mean switching times are well known in the applied math literature for continuous Markov processes described by a Fokker--Planck equation \cite{ludwig75a,talkner87a,maier97a,schuss10a}.
The rigorous mathematical basis of this theory is known as {\em large deviation theory} \cite{freidlin12a,feng06a,kifer09a}.
The theory used to describe metastable behavior for chemical systems generally considers large-$N$-type weak noise conditions \cite{doi76a,peliti85a,hanggi84a,dykman94a,aurell02a,assaf06a}.
The bistable switch has been analyzed using a variety of means to eliminate promotor switching from the problem, by using a diffusion approximation \cite{aurell02a,walczak05a,feng14a}, by taking the adiabatic limit \cite{aurell02a}, or by assuming that mRNA is synthesized in bursts \cite{roma05a}.
However, the first two approaches result in quantitatively inaccurate estimates for the mean switching times \cite{newby13a}, and the latter is only applicable when the mRNA degradation rate is large compared to the transcription rate and the promotor transition rates.

 The first to make progress on developing a general asymptotic approximation, Assaf and coworkers obtained a partial description of bistable switching in a three-species stochastic model (promotor, mRNA, and protein) \cite{assaf11a}.
The result was significant because their model explicitly included mRNA copy number and stochastic ``on-off'' promotor switching.
However, their result does not account for more than two promotor states, and they assumed that mRNA are present in sufficient numbers that it can be treated as a continuous quantity.
Additionally, they derived a logarithmically-accurate asymptotic estimate of the mean switching times, lacking a pre exponential factor (PEF).
Methods for computing the PEF are well developed for the Fokker--Planck equation \cite{ludwig75a,talkner87a,maier97a,schuss10a}, but they have not been widely applied to chemical systems.

We argue that a different approach is necessary to solve the problem, one that applies to chemical systems where weak noise arises from species that can be either ``large $N$'' or ``adiabatic''.
Using theory first developed to study metastability in a molecular motor model with an adiabatic motor configuration \cite{newby11b}, the authors later derived an approximation to the gene expression problem that accounts for an arbitrary number of promotor states \cite{newby12a, newby13a} and a mean switching time approximation that included the PEF, but did not explicitly include mRNA.

In this paper, we develop a complete description of bistable switching in a simple gene regulation circuit that includes promotor switching, a discrete mRNA reaction, and a protein concentration that regulates the promotor switching rates.
Our main assumption is that all of the transition rates (the promotor switching rates, the mRNA transcription and degradation rates, and the protein synthesis rate) are large compared to the protein degradation rate.
Physically, this assumption is valid in a given system if (i) protein is present in sufficient quantity that it can be regarded as a concentration, (ii) mRNA is present in small number, and (iii) intrinsic noise weakly affects the protein concentration.

Using a recently developed quasi-stationary analysis (QSA) \cite{newby13a}, we obtain a Arrhenius--Eyring--Kramers rate that includes the previously unknown PEF.
Our result agrees with the logarithmically accurate approximation reported in \cite{assaf11a} under a less restrictive set of assumptions (we make no assumption about the rate of transcription compared to the rate of mrNA degradation).
In addition to the Kramers rate, the PEF allows us to derive a uniformly accurate asymptotic approximation of the joint stationary probability distribution, including the discrete conditional distribution of mRNA.
The theory is independent of the particular choice of protein dependent promotor switching rates.

The paper is organized as follows. 
First, we introduce the model in Section \ref{sec:model}, along with the deterministic limit.
In Section \ref{sec:quasi-stat-appr} we introduce the QSA and the approximation formula for mean switching times.
The WKB approximation of the stationary probability density function is calculated in Section \ref{sec:wkb}.
Finally, in Section \ref{sec:results} we compare our results with Monte-Carlo simulations (obtained using the standard Gillespie algorithm) for a simple example of positive feedback regulation.

\section{Model}
\label{sec:model}
Let $s$ represent the gene state with $s = 1$ when the gene is on and $s = 0$ when it is off.
When the gene is on, {mRNA} $M$ is transcribed at a rate $\sigma/\epsilon$,  and each mRNA is removed at a constant rate $\gamma/\epsilon$.
Assume that the transitions are fast so that $\epsilon \ll 1$ is a small parameter.
Each mRNA synthesizes protein $X$ at a rate $k_{d}/\epsilon$ and each protein molecule is removed at a rate $k$.
Then, we have the following set of chemical reactions,
\begin{align*}
  \emptyset &\chem{s\sigma/ \epsilon}{\gamma / \epsilon} M \\
  M &{{{k_{d} / \epsilon}\atop\longrightarrow} \atop} M + X \\
  X &{{{k}\atop\longrightarrow} \atop} \emptyset.
\end{align*}
Set the characteristic time to the average lifetime of a single mRNA so that $\gamma = 1$.
Then, $\sigma$ is the average number of mRNA, assuming the gene is permanently switched on.

Let $n$ be the number of proteins of type $X$, and define the ``concentration'' of $X$ to be $x = \epsilon n$.
Note that $x$ is not a physical concentration since $\epsilon$ is a non dimensional parameter.
Assume that $X$ regulates the gene activity by affecting the promotor switching rates.
The gene switches off ($s=0$) and on ($s=1$) randomly according to the two state Markov process
\begin{equation}
  \label{eq:5}
  (\mbox{off})\chem{\alpha(x) / \epsilon}{\beta(x) / \epsilon}(\mbox{on}).
\end{equation}
The analysis presented here is independent of the particular choice of $\alpha(x)$ and $\beta(x)$.

The master equation for the process is
  \begin{equation}
  \label{eq:6}
    \pd{}{t}\pr(s, m, x, t) =   \frac{1}{\epsilon}\left[\mathbb{L}^{(s)} + \mathbb{L}^{(m)}\right]\pr + \mathbb{L}^{(x)}\pr,
  \end{equation}
where
\begin{align}
  \label{eq:1}
  \mathbb{L}^{(s)}[f](s) &\equiv (2s - 1)(\alpha f(0) - \beta f(1)), \\
  \mathbb{L}^{(m)}[f](m) &\equiv s \sigma [f(m-1) - f(m)] \\ \nonumber
  &\qquad + \gamma[(m+1)f(m+1) - mf(m)], \\
  \mathbb{L}^{(x)}[f](x) &=\frac{1}{\epsilon}\left[m k_{d} (e^{-\partial x} - 1)f + k (e^{\partial x} - 1)xf \right].
\end{align}
Formally, we write jump operators $e^{\pm\partial x}$ in terms of a Taylor's series expansion with
\begin{equation}
  \label{eq:27}
  e^{\pm \partial x}f(x) \equiv \sum_{j = 0}^{\infty}\frac{(\pm \epsilon)^{j}}{j!}\frac{d^{j}}{dx^{j}}f(x) = f(x \pm \epsilon).
\end{equation}

\subsection{Deterministic dynamics}
\label{sec:determ-dynam}
In the limit $\epsilon \to 0$, the proceses becomes deterministic, with
\begin{equation*}
  s \to \varphi_{\rm on}(x)\equiv \frac{\alpha(x)}{\alpha(x) + \beta(x)}, \quad m \to \sigma \varphi_{\rm on}(x).
\end{equation*}
The concentration of protein satisfies
\begin{equation}
  \label{eq:34}
  \frac{dx}{dt} = V(x) \equiv k_{d} \sigma \varphi_{\rm on}(x)  - kx.
\end{equation}
Assume that \eqref{eq:34} is bistable for a range of parameter values, having three fixed points, two of which are stable.
Label the two stable fixed points $x_{\pm}$ and the unstable fixed point $x_{*}$ so that $0 < x_{-}<x_{*}<x_{+}$.
For a discusion on how the choice of $\alpha$ and $\beta$ affect stability see Ref.~\cite{warren05a}.

\section{Quasi-stationary analysis}
\label{sec:quasi-stat-appr}
The master equation \eqref{eq:6} can be written as
\begin{equation}
  \label{eq:45}
  \pd{}{t}\pr(s, m, x, t) =  \mathcal{L}_{\epsilon}\pr, 
\end{equation}
where we have defined the linear operator
\begin{equation}
  \label{eq:18}
  \mathcal{L}_{\epsilon} \equiv \frac{1}{\epsilon}\left[\mathbb{L}^{(s)} + \mathbb{L}^{(m)}\right] + \mathbb{L}^{(x)}.
\end{equation}
The solution to \eqref{eq:45} can be written in terms of the eigenvalues $\lambda_{j}$ and eigenfunctions $\efn{j}$ of $-\mathcal{L}_{\epsilon}$ with
\begin{equation}
  \label{eq:52}
  \pr(s, m, x, t) = \sum_{j = 0}^{\infty}c_{j}\efn{j}(s, m, x)e^{-\lambda_{j}t}.
\end{equation}

The process looks very different depending on whether it starts at $x_{0}<x_{*}$ or at $x_{0}>x_{*}$.  
For the sake of illustration assume that $x_{0}=x_{-}$.  
On intermediate time scales, the solution will converge to a stationary density around $x_{-}$ that, figuratively speaking, does not see beyond $x_{*}$ to the other stable fixed point.
Slowly, over a long timescale, the solution converges to the full stationary density as probability slowly leaks out past $x_{*}$ toward $x_{+}$.  
The timescale for this long-time convergence is exponentially large (i.e., $O(e^{C^{2}/\epsilon}$)).  
Since a stationary solution exists, the smallest eigenvalue $\lambda_{0}$, called the principal eigenvalue, is $\lambda_{0}=0$, and the stationary density is the eigenfunction $\efn{0}(s, n, x)$ (up to a normalization constant).

The separation of time scales in the problem can be exploited to approximate the solution.
To understand how this works consider the process where a boundary condition is placed at $x_{*}$ so that the process truly does not see beyond the unstable fixed point.
We want to consider two different boundary conditions: reflecting and absorbing.  
To distinguish between each case, we write the principal eigenvalue and eigenfunction (dropping the subscript) as $\eigabs{},\;\efnabs{}$ and $\eigref{},\;\efnref{}$ for absorbing and reflecting boundary conditions, respectively.
If we place a reflecting boundary at $x_{*}$ the principal eigenvalue $\eigref{}=0$, but the eigenfunction $\efnref{}$ is now restricted to $x\in  (-\infty, x_{*})$ (or $x\in  ( x_{*}, \infty)$ if we instead assume that $x_{0}>x_{*}$).
We call $\efnref{}(s, n, x)$ the {\em quasi-stationary density}; it is a solution to
\begin{equation}
  \label{eq:28}
  \mathcal{L}_{\epsilon}\efnref{} = 0.
\end{equation}
Note that $\efnref{}$ is defined up to a normalization factor.
One of the nice things about the quasi-stationary density is that it can be approximated using the Wentzel--Kramers--Brillouin (WKB) method.

Now suppose that an absorbing boundary is imposed at $x_{*}$.
In this case, no stationary density exists, and the principal eigenvalue is perturbed by an exponentially small amount, that is, $\eigabs{}=O(e^{-C/\epsilon})$, for some $C>0$.  
The eigenfunction $\efnabs{}$ is also perturbed, but away from the boundary, $\efnabs{} \sim \efnref{}$.
Thus, if we can calculate the eigenvalue and eigenfunction, we have an accurate approximation to the absorbing boundary problem with
\begin{equation}
  \label{eq:53}
  \pr(s, m, x, t) \sim \mathcal{N}\efnref{}(s, m, x)e^{-\eigabs{} t},
\end{equation}
where $\mathcal{N}$ is a normalization constant.

The quantity we are most interested in calculating is the mean first exit times to switch between $x_{\pm}$.
Let $\tau$ be the first exit time for the process, having started at $x_{-}$, to reach $x_{*}$.
From \eqref{eq:53}, the survival probability is
\begin{equation*}
  \text{Prob}[t < \tau] = \sum_{s, m}\int_{-\infty}^{\infty}\pr(s, m, x) dx \sim e^{-\eigabs{} t}.
\end{equation*}
It follows that the first exit time is approximately an exponential random variable with mean $T = 1/\eigabs{}$.

The quasi-stationary density and the principle eigenvalue are approximated as follows.
The WKB approximation of $\efnref{}$ proceeds with the anzatz,
\begin{multline}
  \label{eq:29}
  \efnref{}(s, m, x) \sim \\ \pfac(x)\left[\isd(s, m | x)  + \epsilon \isd^{(1)}(s, m, x)\right]e^{-\frac{1}{\epsilon} \Phi(x)},
\end{multline}
where $\isd$ is the conditional distribution for the gene/mRNA states and $\Phi$ is called the quasipotential.
The PEF $\pfac(x)$ can be viewed as a normalization factor for $\isd$.

Let us write the principle eigenvalue $\eigabs{}$ corresponding to $x_{0} = x_{\pm}$ as $\lambda_{\pm}$ so that the mean exit time to transition from $x_{\pm} \to x_{*}$ is given by $T_{\pm} = 1/\lambda_{\pm}$.
Using a spectral projection method \cite{newby13a}, one can derive an asymptotic approximation of the principle eigenvalue given by,
\begin{equation}
  \label{eq:2}
\begin{split}
  \lambda_{\pm} &\sim \frac{V'(x_{*})}{\pi \Phi''(x_{*})}\sqrt{\abs{\Phi''(x_{*})}\Phi''(x_{\pm})} \frac{\pfac(x_{*})}{\pfac(x_{\pm})} \\
&\qquad \qquad \times \explr{-\frac{1}{\epsilon}\left(\Phi(x_{*}) - \Phi(x_{\pm})\right)},
\end{split}
\end{equation}
where $V(x)$ is given by \eqref{eq:34}.
The above formula is known in the literature as the Arrhenius--Eyring--Kramers reaction rate formula \cite{hanggi90a}.

In the next section we calculate the WKB approximation, which yields an approximation of the stationary density function and, using \eqref{eq:2}, the mean switching times.

\subsection{WKB approximation}
\label{sec:wkb}
Applying the jump operators $e^{\pm\partial x}$ defined by \eqref{eq:27} to the WKB solution \eqref{eq:29} and expanding in powers of $\epsilon$ involves expressions of the type
\begin{equation}
  \label{eq:12}
  e^{\pm\partial x}[g(x)e^{-\Phi(x)/\epsilon}] \sim \left[g(x)e^{\mp \Phi'(x)} + O(\epsilon)\right]e^{-\Phi(x)/\epsilon},
\end{equation}
where $g(x)$ is an arbitrary function.
Substituting \eqref{eq:29} into \eqref{eq:28} and collecting leading order terms in $\epsilon$ yields
\begin{equation}
  \label{eq:31}
  \left[\mathbb{L}^{(s)} + \mathbb{L}^{(m)} + mk_{d}(e^{ p} - 1) + xk(e^{- p}-1)\right] \isd(s, m | x) = 0,
\end{equation}
where
\begin{equation}
  \label{eq:30}
  p \equiv  \Phi'(x).
\end{equation}
Note that $p = 0$ at $x = x_{fp}$ so that $\Phi(x)$ has local minima/maxima at the deterministic fixed points.
The goal of the first part of this section is to compute $\Phi'(x)$ and $\isd(s, m | x)$ (the PEF is determined at higher order).
It is rarely possible to integrate $\Phi'(x)$ to get a closed form solution for $\Phi(x)$.
However, using Chebyshev interpolation, the solution can be efficiently computed numerically to any desired accuracy.
There are many software packages that compute Chebyshev approximations, including the {\sc GNU Scientific Library}, which can be easily used from within {\sc Python}.
For {\sc Matlab}, the {\sc Chebfun} package provides the necessary tools.

For notational convenience, let
\begin{equation}
  \label{eq:51}
  v(p) \equiv k_{d}(e^{ p} - 1), \quad u(x, p) \equiv -k x(e^{- p}-1).
\end{equation}
We proceed by developing a solution of the associated eigenvalue problem,
\begin{equation}
  \label{eq:16}
  \left[\mathbb{L}^{(s)} + \mathbb{L}^{(m)} + mv(p) - u(x, p) - \mu(x, p) \right] r_{s, m}(x, p) = 0,
\end{equation}
where $\mu(x, p)$ is the eigenvalue and $r_{s, m}(x, p)$ the eigenvector.
Then, $\Phi'(x)$ is implicitly defined by setting $\mu(x, \Phi'(x)) = 0$.
Given $\Phi'(x)$, the conditional distribution $\isd$ is $r_{s, m}(x, \Phi'(x))$ up to a normalization factor.
For the case where the dimension of linear operator in \eqref{eq:16} is finite (i.e., a matrix), it follows from the Perron--Frobenius Theorem that there is a unique eigenvalue called the {\em principal eigenvalue} corresponding to a nonnegative eigenvector.
The principal eigenvector is real, simple, and is greater than the real part of all other eigenvalues.
We assume that the statement holds in the present situation when $v<1$.
Define the generating function,
\begin{equation}
  \label{eq:43}
  G_{s}(z; x, p) \equiv \sum_{m=0}^{\infty}z^{m}r_{s, m}(x, p).
\end{equation}
Multiplying both sides of \eqref{eq:16} by $z^{m}$ and summing over all $m$ yields (in component form),
\begin{align}
  \label{eq:15}
  ((1-v)z - 1) \pd{}{z}G_{0} &= -(u + \mu) G_{0}\\ \nonumber
&\qquad  - \alpha G_{0} + \beta G_{1},  \\
  \label{eq:17}
  ((1-v)z - 1) \pd{}{z}G_{1} &= (\sigma (z-1) - u - \mu) G_{1} \\\nonumber
&\qquad + \alpha G_{0} - \beta G_{1}.
\end{align}
We can transform the above system into a single second order equation.
Rearranging the first equation to obtain $G_{1}$ in terms of $G_{0}$ yields
\begin{equation}
  \label{eq:10}
\beta G_{1} = ((1-v)z - 1) \pd{G_{0}}{z} + (\alpha + u + \mu) G_{0}
\end{equation}
After substituting \eqref{eq:10} into \eqref{eq:17}, changing variables with $t = \frac{\sigma}{(1-v)^{2}}((1-v)z - 1)$, and setting $y(t) = G_{0}(z(t))$ we obtain the second order equation,
\begin{equation}
  \label{eq:54}
  ty'' + (b - t) y'  - \left(a  + \frac{c}{(1-v)^{3}t}  \right) y = 0,
\end{equation}
where
\begin{align}
  \label{eq:44}
  a & \equiv \frac{\alpha + u + \mu}{1 - v}, \\
  b &\equiv  1 + \frac{\alpha + \beta + 2(u + \mu)}{1 - v} - \frac{\sigma v}{(1 - v)^{2}}, \\
  c &\equiv   \sigma(\alpha + u + \mu) \\ \nonumber
  &\quad - (1-v)[(\sigma + u + \mu)(\alpha + u + \mu)  + \beta(u + \mu) ].
\end{align}

Recall that at fixed points, we must have $p = 0$, and notice that $v(0) = u(x, 0) = 0$.
If we set $p = 0$,  $\mu = 0$, and $y = y_{0}$ in \eqref{eq:54} it simplifies to
\begin{equation}
  \label{eq:55}
  ty_{0}'' + (b_{0} - t) y_{0}'   - \alpha  y_{0} = 0,  
\end{equation}
where $b_{0} = 1 + \alpha + \beta$.
The solution is
\begin{equation}
  \label{eq:14}
  y_{0}(t) = F(\alpha, b_{0}, t) \equiv \sum_{n=0}^{\infty}\frac{\Gamma(b_{0})\Gamma(\alpha + n)t^{n}}{\Gamma(\alpha)\Gamma(b_{0} + n)n!},
\end{equation}
where $F$ is the so-called Kummer function or confluent hypergeometric function (sometimes written as ${}_{1}F_{1}$) and $\Gamma$ is the gamma function.
The solution \eqref{eq:14} is consistent with results found in Ref.~\cite{raj06a} for the generating function of the distribution of mRNA transcribed by an on-off gene (i.e., ignoring protein synthesis and regulation).
Similar results utilizing generating function methods that involve $F$ have been obtained for a variety of linear feedback regulation models \cite{hornos05a,visco08a,visco09a,venegas-ortiz11a}.
For $p \neq 0$,  we notice that there is a solution of the form $y(t) = F(a, b, t)$ provided that $c = 0$.
Of course, there are an infinite number of solutions, one for each of the eigenfunctions of the compact infinite dimensional linear operator \eqref{eq:16}.
Assuming there is a unique nonnegative eigenvector (as is the case for appropriately defined finite dimensional matrices), we can confirm that we have selected the correct solution if the inverse transform of $G$ is nonnegative (up to a normalization factor).
Setting $c = 0$ yields the characteristic equation,
\begin{multline}
  \label{eq:13}
    \mu^{2} 
  + \left[\alpha + \beta + 2u - \frac{\sigma v}{1-v}\right]\mu\\
  - \left[(\alpha + u) \frac{\sigma v}{1-v} - (\alpha + \beta) u - u^{2}\right] = 0.
\end{multline}
To obtain the WKB solution, we must solve for $p(x)$ satisfying $\mu(x, p) = 0$.
Substituting $\mu = 0$ into \eqref{eq:13} yields
\begin{equation}
  \label{eq:24}
  v\left[u^{2} + (\alpha + \beta + \sigma)u + \sigma \alpha \right] - (\alpha + \beta)u - u^{2} = 0.
\end{equation}
Let $\xi = e^{- p}$ and rewrite \eqref{eq:51} as $v(\xi) = k_{d}(1/\xi - 1)$ and $u(x, \xi) = -k x(\xi - 1)$.
From the latter we have $\xi = 1 -  u / (kx)$, which we substitute into $v(\xi)$ to get
\begin{equation}
  \label{eq:26}
  v = \frac{k_{d} u}{k x -  u}.
\end{equation}
After substituting \eqref{eq:26} into \eqref{eq:24}, we find that $u$ is a root of
\begin{equation*}
  u^{2} + \left(\alpha + \beta + \frac{k_{d}\sigma - kx}{k_{d} + 1}\right)u +  \frac{k_{d}\sigma \alpha -  kx(\alpha + \beta)}{k_{d} + 1} = 0,
\end{equation*}
There is one root that vanishes when $x = x_{fp}$, namely
\begin{align}
  \label{eq:33}
    u(x) &= -W(x) + \sqrt{W(x)^{2} - \frac{(\alpha(x) + \beta(x))V(x)}{k_{d} + 1}}, \\
    W(x) &\equiv \frac{1}{2}\left[\frac{k_{d}\sigma - kx}{k_{d} + 1} + \alpha(x) + \beta(x) \right],
\end{align}
where $V(x)$ is the deterministic dynamics \eqref{eq:34} (for which $V(x_{fp}) = 0$ by definition).
Then, using \eqref{eq:51} we obtain,
\begin{equation}
  \label{eq:20}
  \Phi'(x) = - \ln(1 - \frac{u(x)}{k x}).
\end{equation}
Interestingly, \eqref{eq:20} has the same form as the equivalent expression in Ref.~\cite{assaf11a}, which was derived under a stricter set of assumptions.
Because the WKB method is more commonly applied to large-$N$-type weak noise conditions, they made the initial assumption that mRNA can be treated as a concentration (i.e., that $\gamma \ll \sigma$).
Later in the analysis, after the WKB expansion, they use a fast slow analysis to obtain $\Phi'$ by assuming that the rate of mRNA degradation is much larger than the transcription rate (i.e., $\gamma \gg \sigma$), seemingly at odds with their initial assumption.
The derivation of \eqref{eq:20} makes no assumption about the size of $\gamma$ relative to $\sigma$.

Now that $\Phi'$ has been determined, the conditional distribution $\isd$ is
\begin{equation}
  \label{eq:56}
  \isd(s, m | x) = A(x)r_{s, m}(x, \Phi'(x)),
\end{equation}
where $A$ is a normalization factor given by
\begin{equation}
  \label{eq:57}
  A(x) \equiv\sum_{s = 0, 1}\sum_{m=0}^{\infty}r_{s, m}(x, \Phi'(x)) = \lim_{z\to1}\sum_{s = 0, 1}G_{s}(z; x, \Phi'(x)).
\end{equation}
Hence, to determine $\isd$ we need the right eigenvector $r$ and its generating function $G_{s}$.
Setting $p = \Phi'(x)$ allows us to solve \eqref{eq:54} and obtain,
\begin{equation}
  \label{eq:42}
  G_{0}(z; x, \Phi'(x)) = F(a(x), b(x), \frac{\sigma((1 - v(x))z - 1)}{(1 - v(x))^{2}}),
\end{equation}
where $a(x)$ and $b(x)$ are given by \eqref{eq:44} (with $\mu = 0$, $v = v(x)$, and $u = u(x)$).
Recall that $v(x) = \frac{k_{d} u(x)}{k x - u(x)}$ and $u(x)$ is given by \eqref{eq:33}.
The generating function $G_{1}$ is written in terms of $G_{0}$ using \eqref{eq:10}.

We recover $r$ from the generating function using the inverse transform,
\begin{equation*}
  r_{0, m}(x, \Phi'(x)) = \lim_{z\to 0}\frac{\partial^{m}G_{0}(z; x, \Phi'(x))}{\partial z^{m}}.
\end{equation*}
After some calculation, we obtain
\begin{widetext}
\begin{align}
  \label{eq:23}
  \isd(0, m | x)
 & = \frac{\Gamma(b)\Gamma(a + m)}{A\Gamma(a)\Gamma(b + m)}
\left(\frac{\sigma}{1 - v}\right)^{m}\frac{e^{-\frac{\sigma}{(1 - v)^{2}}}}{m!}
F(b - a, b + m, \frac{\sigma}{(1 - v)^{2}}), \\
  \isd(1, m | x) &= \frac{1}{\beta(x)}\left[(\alpha(x) + m(1 - v(x)) + u(x))\isd(0, m | x) - (m+1)\isd(0, m + 1| x)\right],
\end{align}
\end{widetext}
where $F$ is defined by \eqref{eq:14}.
In practice, the validity of the approximation can be verified by confirming that the above distribution is nonnegative.
A general proof of this based on precise assumptions about the model parameters is beyond the scope of this paper.
However, it follows immediately that if $0< a(x) < b(x)$ then $\isd(0, m | x)\geq 0$.
We anticipate that this is true when all the parameters ($x$, $k$, $k_{d}$, $\sigma$, $\alpha$, and $\beta$) are positive.

\subsubsection{Pre exponential factor}
\label{sec:prefactor}
Collecting $O(\epsilon)$ terms in the WKB expansion yields
\begin{equation}
\label{eq:36}
\begin{split}
  &[\mathbb{L}^{(s)} + \mathbb{L}^{(m)} + mv - u]\isd^{(1)}(s, m, x)\\
    & \quad = \pfac(x)\left[-\pcd{u}{p}{x} + \frac{1}{2}\Phi''(x)\left(m \pdd{v}{p} - \pdd{u}{p} \right)  \right]\isd \\
    &\qquad+ \left(m\pd{v}{p} - \pd{u}{p}\right)\left(\pfac'(x)\isd + \pfac(x)\pd{\isd}{x}\right),
 \end{split}
\end{equation}
where $u(p)$ and $v(x, p)$, defined by \eqref{eq:51}, and their derivatives are evaluated at $p = \Phi'(x)$, given by \eqref{eq:20}.
Recall that $v(\Phi'(x))$ and $u(x, \Phi'(x))$ are given by \eqref{eq:26} and \eqref{eq:33}.
More details on obtaining the above expression (namely the second order term in \eqref{eq:12}) can be found in Ref.~\cite{newby13a}.
The PEF is determined by a solvability condition, which makes use of the left eigenvector,
\begin{align}
  \label{eq:39}
  l_{s, m} &\equiv C_{s}\zeta^{m}, \\
 C_{s}(x) &= \frac{\alpha(x) + su(x)}{\alpha(x) + (u(x) + \alpha(x))^{2}/\beta(x)}, \\
   \zeta(x) &= \frac{1}{1 - v(x)}.
\end{align}
The derivation can be found in Appendix \ref{sec:adjoint-problem}.
Define the inner product according to
\begin{equation}
  \ave{a, b} \equiv \sum_{s = 0,1}\sum_{m = 0}^{\infty}a(s, m)b(s, m).
\end{equation}
It follows from the Fredholm Alternative Theorem \cite{keener00c} that a solution $\isd^{(1)}$ to \eqref{eq:36} exists provided that
\begin{multline}
  \label{eq:37}
      \pfac'\left(\pd{u}{p}\ave{l, \isd} - v'\ave{l, m\isd} \right) \\
=  \pfac\left(v'\ave{l, m\pd{\isd}{x}} - \pd{u}{p}\ave{l, \pd{\isd}{x}}\right)\\
   + \pfac\left[-\pcd{u}{p}{x} \ave{l, \isd} + \frac{1}{2}\Phi''\left(v''\ave{l, m\isd} - \pdd{u}{p}\ave{l, \isd} \right)\right].  
\end{multline}

The inner products can be evaluated explicitly using the generating function for $r_{s, m}(x, \Phi'(x))$.
It is simpler to use the unnormalized eigenvector to evaluate the inner products.
Recall that $r_{s, m}(x, \Phi'(x)) = A(x)\isd(s, m | x)$, where $A$ is a normalization factor defined by \eqref{eq:57}.
Hence, $\ave{l, \isd} = A\ave{l, r}$.

Using the generating function $G_{s}(z ; x, p)$, given by \eqref{eq:42} and \eqref{eq:10}, the inner product of the left and right eigenvector is
\begin{equation}
\begin{split}
  Z \equiv \ave{l, r} &=  \sum_{s=0,1}C_{s}\sum_{m = 0}^{\infty} \zeta^{m}r_{s, m} \\
& = \lim_{z\to \zeta}\sum_{s=0,1}C_{s}G_{s}(z; x, \Phi'(x)).
\end{split}
\end{equation}
Note that we have normalized $l_{s, m}$ so that $Z = 1$.
Likewise, we define
\begin{align}
  \widehat{Z}(x) & \equiv \ave{m l, r}  = \zeta(x)\sum_{s=0,1}C_{s}(x)\pd{}{z}U_{s}(\zeta(x); x),  \\
  \widehat{Z}_{x}(x) &\equiv \ave{ml, \frac{dr}{dx}} = \zeta(x)\sum_{s=0,1}C_{s}(x)\pcd{}{z}{x}U_{s}(\zeta(x); x), \\
  Z_{x}(x) &\equiv \ave{l, \frac{dr}{dx}}  = \sum_{s=0,1}C_{s}(x)\pd{}{x}U_{s}(\zeta(x); x),
\end{align}
where $U_{s}(z; x) \equiv G_{s}(z; x, \Phi'(x))$.
The various partial derivatives of the generating function simplify considerably when evaluated at $z = \zeta$; they are listed in Appendix \ref{sec:deriv-gener-funct}.
With the above inner products, we can write the PEF as
\begin{align}
  \label{eq:32}
  \pfac(x) &=  A(x)e^{ - \Psi(x)}, \\
  \Psi'(x) &= \frac{H_{px}(x) + \frac{1}{2}\Phi''(x)H_{pp}(x)}{H_{p}(x)},
\end{align}
where
\begin{align}
  H_{p}(x)  &\equiv \pd{v}{p}\widehat{Z}(x) - \pd{u}{p}, \\
  H_{px}(x) &\equiv \pd{v}{p}\widehat{Z}_{x}(x) - \pd{u}{p}Z_{x}(x),\\
  H_{pp}(x) &\equiv  \pdd{v}{p}\widehat{Z}(x) - \pdd{u}{p}.
\end{align}
The partial derivatives of $u$ and $v$ are evaluated at $p = \Phi'(x)$, which is given by \eqref{eq:20}.
Note that $\Psi'(x)$ contains removable singularities at the fixed points, and is best evaluated using a Chebyshev approximation.

\section{Results}
\label{sec:results}
Suppose that there is a background concentration of active inhibitor $R$ that binds to the DNA and turns the gene off.
Suppose further that the protein $X$ deactivates the inhibitor through the reaction,
\begin{equation}
  \label{eq:48}
  2X + R \chem{}{} R^{*},
\end{equation}
where $R^{*}$ is the deactivated inhibitor.
Assuming that this reaction is fast, a simple way to include regulation in the model is to set
\begin{equation}
  \label{eq:19}
  \beta(X) =  \frac{\beta_{0}}{1 + \kappa X^{2}},
\end{equation}
where $\beta_{0}$ and $\kappa$ are positive parameters.

We compute the WKB \eqref{eq:29} and mean switching time \eqref{eq:2} approximations in {\sc Python} using the {\sc Scipy} package for plotting.
We numerically integrate $\Phi'(x)$ \eqref{eq:20} and $\Psi'(x)$ \eqref{eq:32} using the Chebyshev approximation toolbox from the {\sc GNU Scientific Library}.
All figure are generated using $100$ interpolation points on the interval $(0, 5.2)$.

\begin{figure*}[htbp]
  \centering
  \includegraphics[width = 14cm]{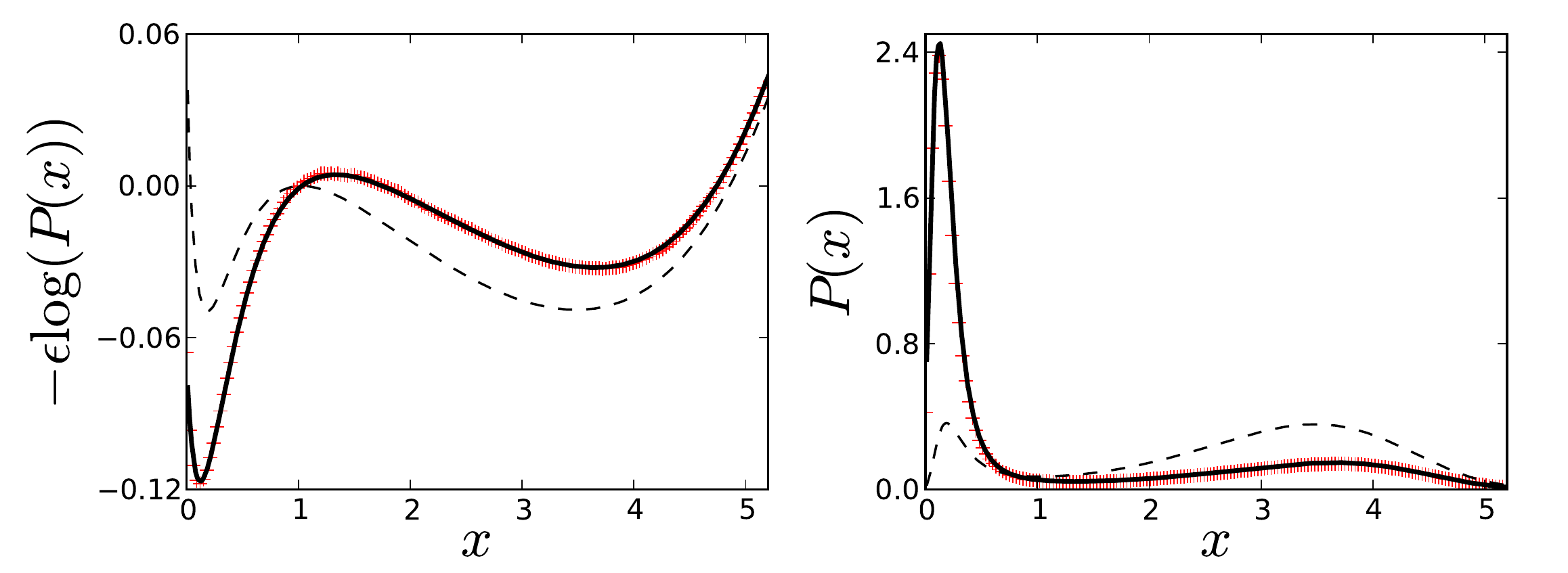}
  \caption{The WKB approximation of the marginal stationary density function $P(x)$ compared to Monte-Carlo simulations (symbols) with $10^{8}$ samples. The solid line shows the approximation with the pre exponential factor and the dashed line show it without. Parameter values are $\alpha = 0.1$, $\beta_{0} = 3$, $\kappa = 7$, $\sigma = 4.7$, $k_{d} = k = 0.3$, and $\epsilon = 0.03$.}
  \label{fig:stat}
\end{figure*}

\begin{figure*}[htbp]
  \centering
  \includegraphics[width=14cm]{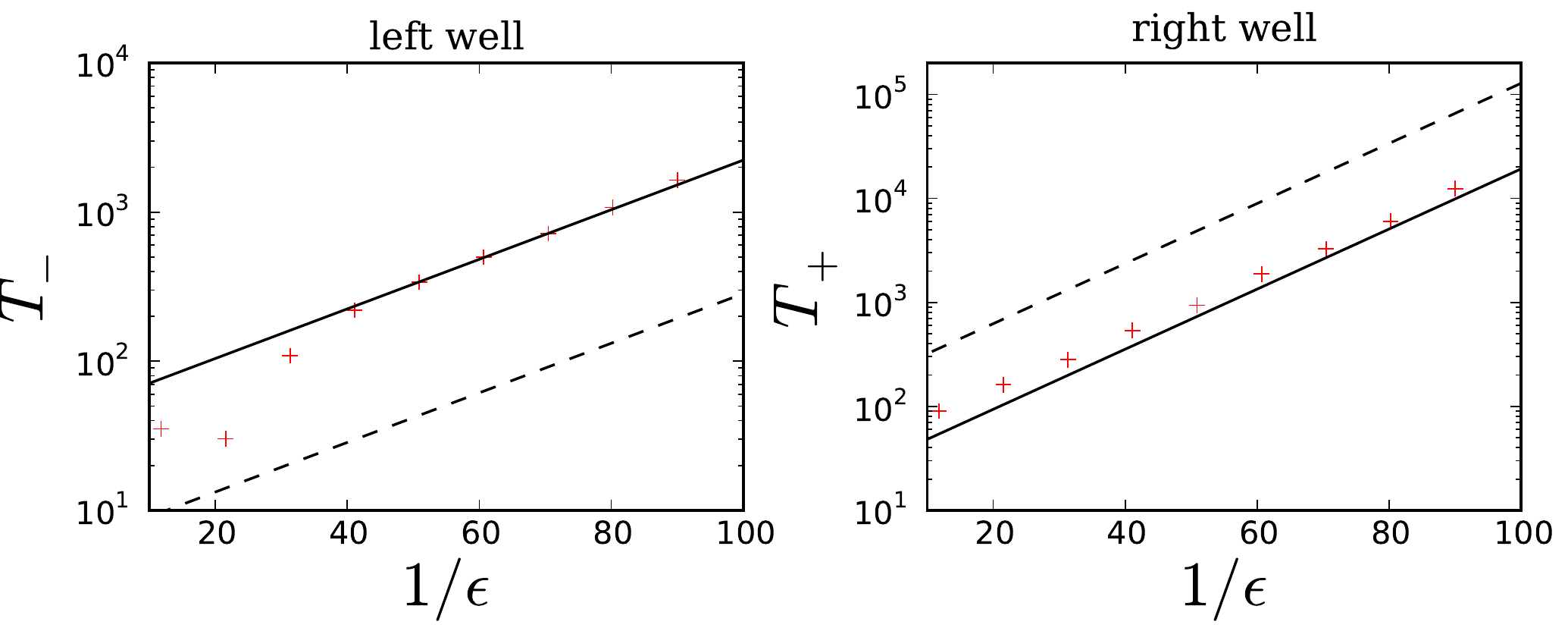}
  \caption{The mean exit time $T_{-}$ from $x_{-}$ to $x_{*}$ and $T_{+}$ from $x_{+}$ to $x_{*}$. Solid line shows the approximation with the prefactor, and the dashed line shows it without. Symbols show $10^{3}$ averaged Monte-Carlo simulations.  Parameter values are the same as in Fig.~\ref{fig:stat}.}
  \label{fig:mfpt}
\end{figure*}

In Fig.~\ref{fig:stat}, we show the WKB approximation of the marginal stationary density function
\begin{equation*}
 P(x) \equiv \sum_{s = 0, 1}\sum_{m = 0}^{\infty}P(s, m, x) \sim \frac{1}{\mathcal{N}} \pfac(x) e^{-\frac{1}{\epsilon}\Phi(x)},
\end{equation*}
where the normalization factor is
\begin{equation*}
  \mathcal{N} \sim \sum_{j  = \pm}\sqrt{\frac{2\pi \epsilon}{\Phi''(x_{j})}}\pfac(x_{j})e^{-\frac{1}{\epsilon}\Phi(x_{j})}.
\end{equation*}
To see the accuracy in the tails of the distribution, we also show $-\epsilon\log(P(x))$.
In Fig.~\ref{fig:mfpt} we show the mean switching times $T_{\pm} \sim 1/\lambda_{\pm}$ as a function of $1/\epsilon$.
The approximations that ignore the PEF are shown as dashed lines for comparison.

\section{Discussion}
Using the QSA, we develop an accurate approximation of the stationary density function and the mean switching times $T_{\pm}$.
Our only assumption is that the protein degradation rate is small compared to all other rates.
Physically, this corresponds to fast promotor and mRNA dynamics and a relatively large number of proteins.
Our assumptions are valid for many physically relevant parameter regimes, including transcriptional bursting when $\gamma \gg \sigma \gg \alpha, \beta$.

Using the generating function for the right eigenvector, we obtain an analytical formula (up to a numerical integration) for the PEF.
The results from a positive feedback model of regulation show that the contribution from the PEF to the stationary density approximation is most significant for small $x$.
It is no surprise then that the PEF is critical for the accuracy of the mean exit time from the left well surrounding stable fixed point $x_{-}$ to the right well.

{ There are also interesting possibilities for how the asymptotic approximation can be used to construct an efficient simulation algorithm.
For continuous Markov processes, many simulation tools have been developed to study rare events, including importance sampling, which can be used in conjunction with the type of asymptotic approximation developed here to speed up simulation time \cite{dupuis12a}. }

The results are derived independent of how regulation is modeled (how $\alpha$ and $\beta$ depend on $x$).
It should be possible to extend these results to more complicated gene regulation circuits and gene networks.
For example, one might consider additional chemical species that interact with the protein synthesized by the gene.
More possibilities exist for metastable behavior in higher dimensions, and analyzing such systems is possible using a recently derived large deviation principle \cite{newby14b}.

\appendix

\renewcommand{\theequation}{A.\arabic{equation}}
\section{Adjoint problem}
\label{sec:adjoint-problem}

We make use of the left eigenvector satisfying
\begin{gather}
  \label{eq:8}
  \left \{ \left[\mathbb{L}^{(s)} + \mathbb{L}^{(m)}\right]^{*} + mv - u \right \} l_{s, m} = 0, \\
  \label{eq:46}
  \ave{l, r} = 1,
\end{gather}
where 
\begin{multline}
  \label{eq:47}
    \left[\mathbb{L}^{(s)} + \mathbb{L}^{(m)} \right]^{*}l_{s, m} = [s\beta - (1 - s)\alpha](l_{0, m} - l_{1, m}) \\
 + s\sigma (l_{s, m+1} - l_{s, m}) +  m (l_{s, m-1} - l_{s, m}).
\end{multline}
Consider the trial solution 
\begin{equation}
  \label{eq:40}
  l_{s, m} = C_{s}\zeta^{m},
\end{equation}
where $C_{0, 1}$ and $\zeta$ are unknown constants.
First, notice that if we substitute \eqref{eq:40} into \eqref{eq:8} with $\mathbb{L}^{(s)} = 0$, we find that $\zeta = 1/(1-v)$.
With $\mathbb{L}^{(s)}$ nonzero and $\zeta = 1/(1-v)$, substituting \eqref{eq:40} into \eqref{eq:8} yields,
\begin{equation*}
  \bvec{ - u - \alpha & \alpha }{\beta(1-v) & \sigma v - (1-v)(u + \beta)}
  \bvec{C_{0}}{C_{1}} = 0
\end{equation*}
Setting the determinant of the above matrix to zero yields an expression equivalent to the characteristic equation \eqref{eq:24} for the principal eigenvalue, which indicates that we have correctly guessed the left eigenvector we need.
Using the normalization condition \eqref{eq:46}, we have
\begin{equation*}
  \bvec{ - u - \alpha & \alpha }{G_{0}(\zeta) & G_{1}(\zeta)}
  \bvec{C_{0}}{C_{1}} = \bvec{0}{1}.
\end{equation*}

\renewcommand{\theequation}{B.\arabic{equation}}
\section{Derivatives of the generation function}
\label{sec:deriv-gener-funct}

Let $h(x) \equiv 1-\frac{k_{d} u(x)}{k x - u(x)}$, with $u(x)$ given by \eqref{eq:33}.
The generating function $G_{s}$ is given by \eqref{eq:42} and \eqref{eq:10}.
Define $U_{s}(z; x) = G_{s}(z; x, \Phi'(x))$ and $Q(t, x) = F(a(x), b(x), t)$, where $a$ and $b$ are given by \eqref{eq:44} and $F$ is defined by \eqref{eq:14}.
For ease of notation, we write partial derivatives of $Q$ with a subscript:

\begin{align}
  \label{eq:59}
  Q_{t}(0, x) &= \frac{(\alpha + u)h}{h^{2} + (\alpha + \beta + 2u)h - \sigma (1-h)}, \\
  Q_{tt}(0, x) &= \frac{(\alpha + u + h)hQ_{t}}{2h^{2} + (\alpha + \beta + 2u)h - \sigma (1-h)}.
\end{align}

Then, $U_{0}(z; x) = Q(\frac{\sigma (hz - 1)}{h^{2}}, x)$.
The $z$ derivatives evaluated at $z = \zeta$ are
\begin{equation}
  \pd{U_{0}}{z} = \frac{\sigma }{h}Q_{t},  \quad
  \pd{U_{1}}{z} = \frac{\sigma}{\beta} \left(1 + \frac{u + \alpha}{h}\right)Q_{t}.
\end{equation}
The $x$ derivatives evaluated at $z = \zeta$ are
\begin{equation}
  \pd{U_{0}}{x}    = \frac{\sigma h'}{h^{3}}Q_{t}, \quad
   \pd{U_{1}}{x}  =\frac{u'}{\beta} + \frac{h'}{h^{2}}\pd{U_{1}}{z}.
\end{equation}
The $z$, $x$ derivatives evaluated at $z = \zeta$ are
\begin{align}
    \pcd{U_{0}}{x}{z} 
    &= \frac{\sigma^{2}h'}{h^{4}}Q_{tt}
+ \frac{\sigma}{h}Q_{tx}
- \frac{\sigma h'}{h^{2}}Q_{t},  \\
\begin{split}
 \pcd{ U_{1}}{x}{z} &= -\frac{\sigma^{2}h'}{\beta h^{4}}(u + \alpha) Q_{tt}
 + \frac{\sigma}{\beta}\left(1 + \frac{u + \alpha}{h}\right) Q_{tx} \\
&\qquad + \frac{\sigma }{\beta h}\left(u' - \frac{h'}{h}(u + \alpha)\right)Q_{t}.
\end{split}
\end{align}

\end{document}